\documentclass[12pt]{article}
\usepackage{indentfirst} 
\usepackage{amsfonts,amssymb,amsmath}
\usepackage{bm}          
\usepackage{cite}
\usepackage{url}
\usepackage{enumitem}
\usepackage[usenames,dvipsnames]{xcolor}
\usepackage{todonotes}
\usepackage{multicol}
\usepackage{MnSymbol}
\usepackage{hyperref}
\usepackage{tikz-cd}
\usepackage[titletoc,title]{appendix}

\usetikzlibrary{arrows}

\AtBeginDocument{}%

\newcommand{\corurl}{red}
\newcommand{\corcite}{ForestGreen}
\newcommand{\corlink}{blue}

\hypersetup{linktocpage,colorlinks,urlcolor=\corurl,citecolor=\corcite,linkcolor=\corlink,
pdftitle={Hamiltonian description of the parametrized electromagnetic field},
pdfauthor={J.F. Barbero, J. Margalef-Bentabol, E.J.S. Villaseñor}}

\numberwithin{equation}{section}  
\begin{document}


\bibliographystyle{plain}

\title{Hamiltonian dynamics of the parametrized electromagnetic field}

\author{
  {\small J. Fernando Barbero G.${}^{1,3}$, Juan Margalef-Bentabol${}^{1,2}$, and
          Eduardo J.S. Villase\~nor${}^{2,3}$} \\[4mm]
  {\small\it ${}^1$Instituto de Estructura de la Materia, CSIC} \\[-0.2cm]
  {\small\it Serrano 123, 28006 Madrid, Spain}         \\[1mm]
  {\small\it ${}^2$Grupo de Modelizaci\'on, Simulaci\'on Num\'erica
                   y Matem\'atica Industrial}  \\[-0.2cm]
  {\small\it Universidad Carlos III de Madrid} \\[-0.2cm]
  {\small\it Avda.\  de la Universidad 30, 28911 Legan\'es, Spain}            \\[1mm]
  {\small\it ${}^3$Grupo de Teor\'{\i}as de Campos y F\'{\i}sica
             Estad\'{\i}stica}\\[-2mm]
  {\small\it Instituto Gregorio Mill\'an, Universidad Carlos III de
             Madrid}\\[-2mm]
  {\small\it Unidad Asociada al Instituto de Estructura de la Materia, CSIC}
             \\[-2mm]
  {\small\it Madrid, Spain}           \\[-2mm]
  {\protect\makebox[5in]{\quad}}  
  \\
}
\date{November 3, 2015}
\maketitle
\thispagestyle{empty}   

\begin{abstract}
We study the Hamiltonian formulation for a parametrized electromagnetic field with the purpose of clarifying the interplay between parametrization and gauge symmetries. We use a geometric approach which is tailor-made for theories where embeddings are part of the dynamical variables. Our point of view is global and coordinate free. The most important result of the paper is the identification of sectors in the primary constraint submanifold in the phase space of the model where the number of independent components of the Hamiltonian vector fields that define the dynamics changes. This explains the non-trivial behavior of the system and some of its pathologies.
\end{abstract}


\clearpage

%
%
\section{Introduction}{\label{sec_intro}}

Parametrized field theories provide interesting examples of relatively simple diff-invariant models. For this reason they have been used as a test bed to understand the quantization of general relativity and related theories \cite{Isham1,Isham2,Kuchar,Hajicek,LR} and recently in the context of loop quantum gravity  \cite{LV1,LV2,LV3}.

An interesting issue that comes up in this setting is the interplay between ordinary gauge symmetries and diffeomorphism invariance. In the canonical treatment of general relativity (and also in the study of parametrized field theories) the standard use of projections onto Cauchy surfaces gives rise to the so called Dirac hypersurface deformation algebra \cite{Dirac2} that replaces the algebra of four-dimensional diffeomorphisms. The fact that the former is very hard to quantize lies at the core of many of the difficulties encountered in the quest for a quantum theory of gravity (see, for example, \cite{Isham0}).

With the purpose of addressing this issue, Isham and Kucha\v{r} \cite{Isham1,Isham2} proposed an approach that, in the case of the scalar field, led to the recovery of the full Lie algebra of four dimensional diffeomorphisms in terms of the Poisson brackets of some functions defined in the full phase space (i.e.\ not only on the primary constraint submanifold in phase space). However, it is not straightforward to extend their procedure to parametrized gauge theories (and parametrized general relativity \cite{Isham2}) due to the non trivial role played by gauge symmetries in this framework \cite{KucharStone1,KucharStone2,Rosenbaum}. Quite unexpectedly, in the particular case of the parametrized Maxwell field, the Gauss law plays a special role to recover the Dirac hypersurface deformation algebra in the full phase space of the model as discussed by the authors of \cite{KucharStone1,Rosenbaum}. As emphasized by Torre in \cite{Torre} ``[...]\textit{ the structure of a parametrized gauge theory is rather different from that of a theory without any gauge invariances.} [...] \textit{It is an interesting problem to find a globally valid formulation of parametrized gauge theory, but we shall not do it here}''.

The purpose of this paper is to shed some light on this issue by relying on geometric methods to derive the structure of the Hamiltonian vector fields (defined on a submanifold of the phase space) that describe the evolution of the system. We will use the ideas developed in \cite{parametrizedboundaries} to study the parametrized scalar field in bounded regions. As we will see, parametrized electromagnetism is, to a certain extent, simpler than those scalar models and is, in fact, quite helpful to understand them.

The layout of the paper is the following. After this introduction, we discuss in section \ref{sec_ParametrizedEM} the Hamiltonian formulation for the parametrized electromagnetic field in a spacetime diffeomorphic to $I\times \Sigma$, with $\Sigma$ a closed (compact without boundary) spatial manifold and $I$ an interval of $\mathbb{R}$. Our starting point is a standard variational principle on the configuration space of the model. We then use geometric methods to derive the form of the Hamiltonian vector fields on the primary constraint submanifold (in the spirit of the Gotay-Nester-Hinds (GNH) geometric constraint algorithm \cite{GNH,Gotay,GN,BPV}). We discuss, in particular, the appearance of sectors on this submanifold characterized by different ``ranks'' of the pullback of the canonical symplectic form. We end the paper with conclusions and comments in section \ref{sec_conclusions}. As we will see, the appearance of the sectors is closely related to a bifurcation in the Dirac algorithm traditionally used in the treatment of constrained systems \cite{Dirac1,Dirac2}. We include two short appendices: in appendix \ref{appendix1} we discuss  this bifurcation phenomenon with our methods in a particular finite dimensional example, and in appendix \ref{appendix2} we compile some mathematical results used in the paper.

%
%
\section{The parametrized electromagnetic field}{\label{sec_ParametrizedEM}}

\subsection{Variational setting}

Let us consider a four-dimensional oriented  and time oriented Lorentzian manifold $(M:=[t_1,t_2]\times\Sigma,g)$ where $\Sigma$ is a closed three-manifold and $\{t\}\times \Sigma$ are Cauchy surfaces for $(M,g)$. The  metric $g$ is taken to have signature $(\varepsilon, +,+,+)$, where the parameter $\varepsilon =- 1$ is introduced to allow for a straightforward extension of our results to the Riemannian case. We will use the Penrose abstract index notation throughout the paper with abstract indices $a,b\ldots$ for $\Sigma$ and $\alpha,\beta\ldots$ for $M$.

The Maxwell action is
\begin{equation}
S_{\mathrm{EM}}(A)=\frac{\varepsilon}{4}\int_M g^{\alpha\beta}g^{\gamma\delta}(\mathrm{d}A)_{\alpha\gamma}(\mathrm{d}A)_{\beta\delta}\mathrm{vol}_g\,,
\label{action_non_param}
\end{equation}
where $A$ is the standard one-form potential, $\mathrm{d}A$ its exterior derivative, and $\mathrm{vol}_g$ denotes the $g$-metric volume form. Parametrization is achieved by introducing diffeomorphisms $Z$ of $M$ as dynamical variables, pulling back with them all the objects used to define the action \eqref{action_non_param} (both the \textit{background} metric $g$ and the \textit{dynamical} 1-form potential $A$), and renaming $Q:=Z^*A$. This way we get the parametrized action
\begin{equation}
S(Q,Z)=\frac{\varepsilon}{4}\int_M (Z^*g)^{\alpha\beta}(Z^*g)^{\gamma\delta}(\mathrm{d}Q)_{\alpha\gamma}(\mathrm{d}Q)_{\beta\delta}\mathrm{vol}_{Z^*g}\,.
\label{action_param}
\end{equation}
From this expression it is a simple exercise (see section 3 of \cite{parametrizedboundaries}) to write down the form of the Lagrangian for parametrized electromagnetism in a configuration space $\mathcal{Q}$ with typical elements $(q_\perp,q_a,X)$ consisting of a smooth scalar field $q_\perp\in C^\infty(\Sigma)$, a smooth 1-form $q_a\in\Omega^1(\Sigma)$ and a smooth $g$-spacelike embedding $X\in \mathrm{Emb}_{g\textrm{-s}}(\Sigma,M)$. The Lagrangian $L:\mathcal{D}\subset T\mathcal{Q}\rightarrow\mathbb{R}$ is given by
\begin{align*}
L(\mathbf{v}_{(q_\perp,q_a,X)})=\frac{\varepsilon}{2} \int_\Sigma&\left( \frac{\gamma_X^{ab}}{n_X(V)}\big(v_a-{\textrm{\L}}_{e^*V}q_a-\mathrm{d}(q_\perp n_X(V))_a\big)
 \big(v_b-{\textrm{\L}}_{e^*V}q_b-\mathrm{d}(q_\perp n_X(V))_b\big)\right.
 \\&\phantom{+}+\left.\frac{\varepsilon n_X(V)}{2}\gamma_X^{ab}\gamma_X^{cd}(\mathrm{d}q)_{ac}(\mathrm{d}q)_{bd}\right)\mathrm{vol}_{\gamma_X}\,,
\end{align*}
where $\mathcal{D}:=\{\mathbf{v}\in T\mathcal{Q}:\varepsilon n_X(V)>0\}$ with $\mathbf{v}_{(q_\perp,q_a,X)}:=(q_\perp,q_a,X;v_\perp,v_a,V^\alpha)\in T_{(q_\perp,q_a,X)}\mathcal{Q}$. The embedding dependent objects $\gamma^{ab}_X$, $(e_X)_\alpha^a$ and $(n_X)_\alpha$ are, respectively, the inverse of the 3-metric $\gamma_X:=X^*g$ on $\Sigma$ induced by the embedding $X$, a projection $(e_X)^a_\alpha$ of the tangent map $TX:T\Sigma\rightarrow TM$ given by $(e_X)^a_\alpha:=g_{\alpha\beta}(TX)^\beta_b\gamma_X^{ba}$ and $(n_X)_\alpha=g_{\alpha\beta}n_X^\beta$, where $n_X^\beta$ is the future directed $g$-unit normal vector field over the $X$ map (see \cite{parametrizedboundaries} for details). We denote the Lie derivative along the vector field $(e_X)^a_\alpha V^\alpha$ as ${\textrm{\L}}_{e^*V}$.

If we choose a fiducial volume 3-form $\mathrm{vol}_\Sigma$ such that $\mathrm{vol}_{\gamma_X}=\sqrt{\gamma_X}\mathrm{vol}_\Sigma$, then the fiber derivative $FL:\mathcal{D}\subset T\mathcal{Q}\rightarrow T^*\mathcal{Q}$ is given by

\begin{align*}
&FL(\mathbf{v}_{(q_\perp,q_a,X)})(\mathbf{w}^1_{(q_\perp,q_a,X)})=\int_\Sigma p^\perp w_\perp \mathrm{vol}_{\Sigma}\,,&&\qquad\mathbf{w}^1_{(q_\perp,q_a,X)}:=(q_\perp,q_a,X;w_\perp,0,0) \,,\\
&FL(\mathbf{v}_{(q_\perp,q_a,X)})(\mathbf{w}^2_{(q_\perp,q_a,X)})=\int_\Sigma p^a w_a \mathrm{vol}_{\Sigma}\,,&&\qquad\mathbf{w}^2_{(q_\perp,q_a,X)}:=(q_\perp,q_a,X;0,w_a,0)\,,\\
& FL(\mathbf{v}_{(q_\perp,q_a,X)})(\mathbf{w}^3_{(q_\perp,q_a,X)})=\int_\Sigma  P_\alpha W^\alpha \mathrm{vol}_{\Sigma}\,,&&\qquad\mathbf{w}^3_{(q_\perp,q_a,X)}:=(q_\perp,q_a,X;0,0,W^\alpha)\,,
\end{align*}
with
\begin{align*}
p^\perp&=0\,,\\
p^a&:=\frac{\sqrt{\gamma_X}}{\varepsilon n_X(V)}\gamma_X^{ab}\big(v_b-{\textrm{\L}}_{e^*V}q_b-\mathrm{d}(q_\perp n_X(V))_b \big)\,,\\
P_\alpha&:=\left(-\frac{1}{2\sqrt{\gamma_X}} (\gamma_X)_{ab}p^a p^b+\frac{\varepsilon}{4}\sqrt{\gamma_X}\gamma_X^{ab}\gamma_X^{cd}(\mathrm{d}q)_{ac}(\mathrm{d}q)_{bd}+\varepsilon q_\perp \nabla_a p^a\right)\varepsilon(n_X)_\alpha\\
&\phantom{:=}+(e_X)_\alpha^a\Big(q_a\nabla_b p^b-p^b(\mathrm{d}q)_{ab}\Big)\,.
\end{align*}
As usual momenta can be considered as 3-form-valued tensor fields or, alternatively, as \textit{densities}.

The fiber derivative is not onto due to the existence of constraints, i.e.\ relations among the canonical variables in phase space. They define the primary constraint submanifold which in the present case is given by
\[
\mathcal{M}:=FL(\mathcal{D})=\{p^\perp=0,\, P_\alpha +  \mathcal{H} \varepsilon n_\alpha+  e_\alpha^a \mathcal{H}_a=0_\alpha\}\subset T^*\mathcal{Q}
\]
where
\begin{align*}
\mathcal{H}(q_\perp,q_a,X,p^b)&:=\frac{1}{2\sqrt{\gamma_X}} (\gamma_X)_{ab}p^a p^b-\frac{\varepsilon}{4}\sqrt{\gamma_X}\gamma_X^{ab}\gamma_X^{cd}(\mathrm{d}q)_{ac}(\mathrm{d}q)_{bd}-\varepsilon q_\perp \nabla_a p^a\,,\\
\mathcal{H}_a(q_b,p^c)&:=p^b(\mathrm{d}q)_{ab}-q_a\nabla_b p^b\,.
\end{align*}
As the Lagrangian is homogeneous of degree one in the velocities, it is straightforward to conclude (by using Euler's theorem on homogeneous functions) that the Hamiltonian on the primary constraint hypersurface is zero. This, of course, can be explicitly checked by computing the Hamiltonian $H$ according to the standard definition (see, for instance, \cite{parametrizedboundaries}). In this situation the complete determination of the dynamics of the system just amounts to solving the equation
\begin{equation}\label{eq hamiltonian}
\imath_Z\omega=\mathrm{d}H=0\,,
\end{equation}
for vector fields $Z$ on $\mathcal{M}$. Here $\omega$ is the pullback to $\mathcal{M}$ of the canonical symplectic form $\Omega$ of $T^*\mathcal{Q}$. Notice that equation \eqref{eq hamiltonian} is equivalent to the determination of the degenerate directions of $\omega$.

We will write vectors in the tangent spaces to $T^*\mathcal{Q}$ at points $(q_\perp,q_a,X;p^\perp,p^a,P_\alpha)$ as
$Y,Z\in T_{(q_\perp,q_a,X;p^\perp,p^a,P_\alpha)}T^*\mathcal{Q}$:
\begin{align*}
Y&=((q_\perp,q_a,X;p^\perp,p^a,P_\alpha),(Y_{\textsf{q}_\perp},Y_{\textsf{q}a},Y_{\textsf{X}}^\alpha,Y_{\textsf{p}^\perp},Y_{\textsf{p}}^a,Y_{\textsf{P}\alpha}))\,,\\
Z&=((q_\perp,q_a,X;p^\perp,p^a,P_\alpha),(Z_{\textsf{q}_\perp},Z_{\textsf{q}a},Z_{\textsf{X}}^\alpha,Z_{\textsf{p}^\perp},Z_{\textsf{p}}^a,Z_{\textsf{P}\alpha}))\,.
\end{align*}
The canonical symplectic form on the cotangent bundle $T^*\mathcal{Q}$ acting on pairs of vector fields takes the following form in the present case
\begin{equation}
\Omega(Y,Z)=\int_\Sigma \Big(Z_{\textsf{p}^\perp}Y_{\textsf{q}_\perp}- Z_{\textsf{q}_\perp}Y_{\textsf{p}^\perp}+Z_{\textsf{p}}^a Y_{\textsf{q}a} - Z_{\textsf{q}a} Y_{\textsf{p}}^a  +Z_{\textsf{P}\alpha}Y_{\textsf{X}}^\alpha- Z_{\textsf{X}}^\alpha Y_{\textsf{P}\alpha}\Big) \mathrm{vol}_\Sigma\,.
\label{EQ_simplectica_cotangente}
\end{equation}
Its pullback to the primary constraint submanifold $\mathcal{M}$ can be simply obtained by computing its action on vector fields on $\mathcal{M}$. These are characterized by the conditions
\begin{equation}
0=D_Zp^\perp=Z_{\textsf{p}^\perp}\,,\qquad 0_\alpha=D_Z(P_\alpha +  \mathcal{H}\varepsilon n_\alpha + e_\alpha^a \mathcal{H}_a )=Z_{\textsf{P}\alpha}+D_Z(  \mathcal{H}\varepsilon n_\alpha +  e_\alpha^a \mathcal{H}_a)\,,
\label{var}
\end{equation}
where $D_Z$ denotes the variations along the vector $Z$ (see appendix \ref{appendix2}). Equation \eqref{var} effectively allows us to obtain the component $Z_{\textsf{P}\alpha}$ in terms of the rest of the components, field values and momenta (the detailed form is given in appendix \ref{appendix2}, together with some hints about how to compute it). Once these tangent fields have been characterized it is possible to write the pulled back symplectic form (after a long but straightforward computation) as
\begin{align}
&\hspace*{-1.7cm}\omega_{(q_\perp,q,X;p)}(Y,Z)\label{EQ_simplectica_simpl}\\
\hspace*{3mm}=\int_\Sigma \mathrm{vol}_\Sigma&\left[ \varepsilon\Big( (Z_{\textsf{q}_\perp}\!\!-\!{\textrm{\L}}_{Z_{\textsf{X}}^{\top}}q_\perp \!+\! q_a\nabla^a Z^\perp_{\textsf{X}} )  Y^\perp_{\textsf{X}} \nabla_b p^b-(Y_{\textsf{q}_\perp}\!\!-\!{\textrm{\L}}_{Z_{\textsf{X}}^{\top}}q_\perp \!+ q_a\nabla^a Y^\perp_{\textsf{X}} ) Z^\perp_{\textsf{X}} \nabla_b p^b   \Big)\right.\nonumber \\
&+\Big(Y_{\textsf{q}a}\!\!-\!{\textrm{\L}}_{Y_{\textsf{X}}^{\top}}q_a\!\!-\!\frac{p_a}{\sqrt{\gamma_X}}Y^\perp_{\textsf{X}}\!\!-\!\varepsilon\mathrm{d}(q_\perp Y^\perp_{\textsf{X}})_a\Big)
          \Big(Z_{\textsf{p}}^a-{\textrm{\L}}_{Z_{\textsf{X}}^{\top}}p^a+\varepsilon \sqrt{\gamma_X} \nabla_c(Z^\perp_{\textsf{X}} (\mathrm{d}q)^{ca})\Big)\nonumber\\
&-\Big(Z_{\textsf{q}a}\!\!-\!{\textrm{\L}}_{Z_{\textsf{X}}^{\top}}q_a\!\!-\!\frac{p_a}{\sqrt{\gamma_X}}Z^\perp_{\textsf{X}}\!\!-\!\varepsilon\mathrm{d}(q_\perp Z^\perp_{\textsf{X}})_a\Big)
          \Big(Y_{\textsf{p}}^a-{\textrm{\L}}_{Y_{\textsf{X}}^{\top}}p^a+\varepsilon \sqrt{\gamma_X} \nabla_c(Y^\perp_{\textsf{X}} (\mathrm{d}q)^{ca})\Big)\Big]\,.\nonumber
\end{align}
Here indices are raised and lowered with $\gamma_X^{ab}$ and $(\gamma_X)_{ab}$ respectively and we have introduced the decomposition $Z_{\textsf{X}}^\alpha=Z_{\textsf{X}}^\perp n_X^\alpha+(TX)^\alpha_a Z_{\textsf{X}}^{\top a}$. Finally, remember that the Lie derivative of the vector density $p^a$ is given by ${\textrm{\L}}_{Z^\top}p^a:=\nabla_b(Z^{ \top b}  p^a) -p^b\nabla_b Z^{\top a}$.

%
%
\subsection{Hamiltonian vector fields}

The degenerate directions for $\omega$ are obtained by requiring $\omega_{(q_\perp,q,X;p)}(Y,Z)=0$ for all $Y\in\mathfrak{X}(\mathcal{M})$. A natural way to get the conditions coming from this is by taking in (\ref{EQ_simplectica_simpl}) all the components of the vector field $Y$ but one equal to zero in all the possible ways. By considering $Y_{\textsf{q}a}$, $Y_{\textsf{p}}^a$, $Y_{\textsf{q}_\perp}$ arbitrary and different from zero in turn we obtain
\begin{align}
&Z_{\textsf{q}a}={\textrm{\L}}_{Z_{\textsf{X}}^{\top}}q_a+\frac{p_a}{\sqrt{\gamma_X}}Z^\perp_{\textsf{X}}+\varepsilon\mathrm{d}(q_\perp Z^\perp_{\textsf{X}})_a\,,\label{Zq}\\
&Z_{\textsf{p}}^a={\textrm{\L}}_{Z_{\textsf{X}}^{\top}}p^a-\varepsilon \sqrt{\gamma_X} \nabla_b(Z^\perp_{\textsf{X}} (\mathrm{d}q)^{ba})\,,\label{Zp}\\
&Z_{\textsf{X}}^\perp\nabla_bp^b=0\,,\label{Gauss}
\end{align}
whereas the vanishing of all the components of $Y$ but $Y_{\textsf{X}}^\alpha$ and the use of \eqref{Zq}-\eqref{Gauss} leads to the extra condition
\begin{equation}
\varepsilon\big(Z_{\textsf{q}_\perp}-{\textrm{\L}}_{Z_{\textsf{X}}^{\top}}q_\perp+q_a\nabla^aZ_{\textsf{X}}^\perp\big)\nabla_bp^b=0\,.
\label{Zqperp}
\end{equation}
The conditions \eqref{Zq} and \eqref{Zp} give us the expressions for $Z_{\textsf{p}}^a$ and $Z_{\textsf{q}a}$ at each point of the primary constraint hypersurface $\mathcal{M}$. Conditions \eqref{Gauss} and \eqref{Zqperp} are quite interesting so we discuss them separately.

%
%
\subsection{Sectors}

The phase space submanifold $\mathcal{M}$, whose points are naturally labeled by a scalar field $q_\perp\in C^\infty(\Sigma)$, a 1-form $q_a\in \Omega^1(\Sigma)$ and a vector density $p^a$, can be divided into sectors. One of them corresponds to the situation when the Gauss law $\nabla_ap^a=0$ holds, while the other one is its complement in $\mathcal{M}$.

In the first case, from the conditions \eqref{Zq}-\eqref{Zqperp} we obtain

\noindent
\begin{minipage}{0.55\textwidth}\noindent
\begin{align}
&Z_{\textsf{q}a}={\textrm{\L}}_{Z^\top_{\textsf{X}}}q_a +\frac{p_a}{\sqrt{\gamma_X}} Z^\perp_{\textsf{X}}+\varepsilon\mathrm{d}( q_\perp Z^\perp_{\textsf{X}})_a\,,\\
&Z_{\textsf{p}}^a={\textrm{\L}}_{Z^\top_{\textsf{X}}}p^a - \varepsilon \sqrt{\gamma_X}\nabla_b\big(Z^\perp_{\textsf{X}}(\mathrm{d}q)^{ab}\big)\,,
\end{align}
\end{minipage}
\hfill
\noindent
\begin{minipage}{0.35\textwidth}\noindent
\begin{align}
&Z_{\textsf{q}_\perp}\hphantom{=}\mathrm{arbitrary}\,,\\
&Z_{\textsf{X}}^\perp\hphantom{=}\,\,\mathrm{arbitrary}\,,\\
&Z_{\textsf{X}}^{\top a}\hphantom{=}\mathrm{arbitrary}\,.
\end{align}
\end{minipage}

\mbox{}\vspace*{1.5ex}

From these expressions it is straightforward to identify the gauge symmetries of the parametrized electromagnetism in this sector: three dimensional diffeomorphisms and the usual gauge transformations of electromagnetism that can be read off directly from the explicit expressions of $Z_{\textsf{q}a}$ and $Z_{\textsf{p}}^a$ given above. For a particular choice of $Z_{\textsf{X}}^\perp>0$ (which is possible only in this sector, as a consequence of \eqref{Gauss}) and $Z_{\textsf{X}}^{\top a}$ we can construct a curve of $g$-spacelike embeddings that can be interpreted as a spacetime diffomorphism. If we have two such curves corresponding to two choices of $Z_{\textsf{X}}^\alpha$, that interpolate between two given Cauchy surfaces, and we compute the integral curves of the Hamiltonian vector fields for a given choice of the scalar potential $q_\perp$, the solutions that we obtain are connected by a four-dimensional diffeomorphism. This is how spacetime diffeomorphisms are implemented in this sector.

Notice that both $Z_{\textsf{X}}^\alpha$ and $Z_{\textsf{q}_\perp}$ are completely arbitrary and, hence, we can choose any scalar potential $q_\perp$ (as in standard electromagnetism) and any spacetime foliation. If we take $Z^\perp_\textsf{X} = 1$, $Z^{\top a}_\textsf{X} = 0^a$ and the initial embedding $X_0:\Sigma\to I\times\Sigma$ given by $X_0(\sigma)=(0,\sigma)$, then we recover the standard dynamics given by the Hamiltonian form of the Maxwell equations; other choices give the Hamiltonian formulation of electromagnetism in whichever foliation we want. Finally notice that the dynamics is perfectly consistent because the Hamiltonian vector fields in this sector are tangent to the submanifold defined by the Gauss law $\nabla_ap^a=0$ (see appendix \ref{appendix2})
\[
D_Z(\nabla_ap^a)=\nabla_a Z_{\textsf{p}}^a =(\nabla_a\nabla_bp^b)Z^{\top a}_{\textsf{X}}+(\nabla_aZ^{\top a}_{\textsf{X}})\nabla_bp^b=0\,.\]
Therefore the integral curves passing through a point in this sector never leave it.

An analogous statement is of course true in the second sector. A simple interpretation of the dynamics is available in the subsector in which $\nabla_ap^a\neq0$ everywhere in $\Sigma$, because then \eqref{Gauss} implies that $Z_{\textsf{X}}^\perp=0$ and the Hamiltonian vector field reads simply\vspace*{-1ex}

\noindent
\begin{minipage}{0.4\textwidth}\noindent
	\begin{align}
	&Z_{\textsf{q}a}={\textrm{\L}}_{Z^\top}q_a\,,\\
	&Z_{\textsf{p}}^a={\textrm{\L}}_{Z^\top}p^a\,,\\
	&Z_{\textsf{q}_\perp}={\textrm{\L}}_{Z^\top}q_\perp\,,
	\end{align}
\end{minipage}
\hfill
\noindent
\begin{minipage}{0.4\textwidth}\noindent
	\begin{align}
	&Z_{\textsf{X}}^\perp=0\,,\\
	&Z_{\textsf{X}}^{\top a}\hphantom{=}\mathrm{arbitrary}\,.
	\end{align}
\end{minipage}

\mbox{}\vspace*{1ex}

The fact that $Z_{\textsf{X}}^\perp=0$ means that the dynamics on this subsector is such that the initial Cauchy hypersurface ``never moves forward''. This means, in the language of \cite{Bauer}, that we move within an equivalence class of shapes. The integral curves of the Hamiltonian vector field just correspond to the action of arbitrary 3-dimensional diffeomorphisms on the fields $q_\perp$, $q_a$ and $p^a$ on $\Sigma$  defined by the arbitrary field $Z_{\textsf{X}}^{\top a}\in \mathfrak{X}(\Sigma)$. Dynamics in this case reduces to the action of 3-dimensional diffeomorphisms.

It is interesting to note that the ``number of independent components'' of the Hamiltonian vector field in different sectors is not the same. In finite dimensional analogues of our system, this behaviour is explained by the fact that the rank of the corresponding $\omega$ is not constant on the primary constraint hypersurface (see appendix \ref{appendix1}).

The issues that we have just discussed are related to similar ones that we found in \cite{parametrizedboundaries} but are simpler to interpret, mainly because in that case the presence of a boundary gave rise to an infinite tower of conditions. Besides, such conditions involved all the phase space variables while here the Gauss law, that defines the sectors in the present case, only depends on a single variable: $p^a$.

%
%
\section{Comments and conclusions}\label{sec_conclusions}

In this paper we have followed a simple geometric approach (actually a very simple application of the GNH approach) to get the Hamiltonian vector fields for parametrized electromagnetism. The appearance of the Gauss law as defining sectors in the primary constraint hypersurface may seem strange but it has a known counterpart in the context of the standard Dirac approach to constrained systems: a bifurcation of the standard algorithm in the parlance of \cite{Henneaux}. Instead of giving a general argument here we will discuss Example 4 of section \textbf{1.6.3} of reference \cite{Henneaux} in appendix \ref{appendix1} from the perspective of the geometric methods that we have used in the paper. This example is, actually, a very good toy model for parametrized electromagnetism: different sectors describe very different physics. Indeed, as we have seen in the previous section, in the sector where the Gauss law holds, we have the standard dynamics of electromagnetism for arbitrary foliation whereas the dynamics of the subsector where $\nabla_ap^a\neq0$ everywhere, although well defined, is trivial in some sense (only the action of diffeomorphisms of $\Sigma$) and does not allow us to build solutions to the field equations that move the initial Cauchy hypersurface.

The fact that the number of independent components in the different sectors of the primary constraint hypersurface $\mathcal{M}$ is not the same means that the reduced phase space will have a very complicated topology as the ``dimensionality'' of the equivalence classes defined by gauge orbits changes from point to point in $\mathcal{M}$. This may also signal a difficulty for the covariant phase space approach to Hamiltonian dynamics as some implicit regularity assumptions are usually made in that setting.

Difficulties of a similar nature will show up when attempting the quantization of the system. For instance, the Dirac approach seems to be very difficult to use if one wants to keep all the sectors at the same time because it hinges on representing the full constraint hypersurface with functions defined on the full phase space. Reduced phase space quantization will be hindered by the difficulties, just mentioned, associated with the singular character of the reduced phase space. It is hard to tell if path integral methods will be easier to use.

A possible way to tackle these problems may be to study the symplectic Lagrangian formulation obtained by pulling back the canonical symplectic structure to the tangent bundle of the configuration space with the help of the fiber derivative. We plan to study this approach in the near future.

%
%
\section*{Acknowledgments}

This work has been supported by the Spanish MINECO research grants FIS2012-34379, FIS2014-57387-C3-3-P and the  Consolider-Ingenio 2010 Program CPAN (CSD 2007-00042). Juan Margalef-Bentabol is supported by a ``la Caixa'' fellowship.

%
%
\begin{appendices}
\section{An example of bifurcation}\label{appendix1}

A discussion of the following finite dimensional example from the point of view of the Dirac algorithm appears in \cite[page 39]{Henneaux}. It is used in that reference to explicitly show the phenomenon of bifurcation that we claim is also responsible for the intricacies of the Hamiltonian formulation of the parametrized electromagnetic field.

We show here that our approach provides the same information that the standard Dirac algorithm. Let us take a system described in the phase space $(T^*\mathbb{R}^3,\Omega)$ coordinatized as $(q_1,q_2,q_3;p_1,p_2,p_3)$, endowed with the canonical symplectic form $\Omega=\mathrm{d}q_1\wedge \mathrm{d}p_1+\mathrm{d}q_2\wedge \mathrm{d}p_2+\mathrm{d}q_3\wedge \mathrm{d}p_3$ and with Hamiltonian $H=0$. Let us consider the ``primary submanifold''
\[
\mathcal{M}=\{(q_1,q_2,q_3,p_1,p_2,p_3)\in T^*\mathbb{R}^3 \,:\, p_1=0,\, p_3=-q_1q_2\}\simeq \mathbb{R}^4\,,
\]
that can be obviously coordinatized with a global chart $(q_1,q_2,q_3,p_2)$. The pullback of $\Omega$ to $\mathcal{M}$ is:
\[
\omega=\mathrm{d}q_2\wedge \mathrm{d}p_2-q_1 \mathrm{d}q_3\wedge \mathrm{d}q_2-q_2\mathrm{d}q_3\wedge \mathrm{d}q_1\,.
\]
On $\mathcal{M}$, the Hamiltonian vector fields $Z$ must satisfy $\imath_Z\omega=\mathrm{d}H=0$ which immediately leads to the conditions
\[
Z_{\textsf{q}_2}\mathrm{d}p_2-Z_{\textsf{p}_2}\mathrm{d}q_2-q_1 Z_{\textsf{q}_3}\mathrm{d}q_2+q_1Z_{\textsf{q}_2}\mathrm{d}q_3-q_2Z_{\textsf{q}_3}\mathrm{d}q_1+q_2Z_{\textsf{q}_1}\mathrm{d}q_3=0\,,
\]
that is,\vspace*{-1ex}

\qquad
\begin{minipage}{0.4\textwidth}\noindent
	\begin{align*}
	Z_{\textsf{q}_2}&=0\,,\\
	Z_{\textsf{p}_2}&=-q_1Z_{\textsf{q}_3}\,,
	\end{align*}
\end{minipage}
\begin{minipage}{0.5\textwidth}\noindent
	\begin{align*}
	q_2Z_{\textsf{q}_1}&=0\,,\\
	q_2Z_{\textsf{q}_3}&=0\,.
	\end{align*}
\end{minipage}\hfill

\mbox{}\vspace*{1ex}

We have now two sectors: If the initial data belong to the sector $\{q_2\neq 0\}\cap \mathcal{M}$ we have only one Hamiltonian vector field given by
\[
Z=(Z_{\textsf{q}_1},Z_{\textsf{q}_2},Z_{\textsf{q}_3},Z_{\textsf{p}_2})=(0,0,0,0)\,,
\]
and, hence, the dynamics is trivial. On the other hand, if the initial data belong to the sector $\{q_2=0\}$ there are infinitely many ``Hamiltonian vector fields'' (which, actually, do not define a smooth vector field on the full primary constraint submanifold $\mathcal{M}$ but only when restricted to $\{q_2=0\}\cap \mathcal{M}$). They are given by
\[
Z=(Z_{\textsf{q}_1},Z_{\textsf{q}_2},Z_{\textsf{q}_3},Z_{\textsf{p}_2})=(Z_{\textsf{q}_1},0,Z_{\textsf{q}_3},-q_1Z_{\textsf{q}_3})
\]
with arbitrary $Z_{\textsf{q}_1}$ and $Z_{\textsf{q}_3}$. As $Z_{\textsf{q}_2}=0$, the vector fields $Z$ are tangent to $\{q_2=0\}\cap \mathcal{M}$ and hence their integral curves remain there.

Notice that, in the notation of \cite{Henneaux} and introducing the ``Lagrange multipliers'' $u_1,u_2\in C^\infty(T^*\mathbb{R}^3)$, the total Hamiltonian defined on $T^*\mathbb{R}^3$ would have the form \[
H_T=u_1 p_1+u_2(p_3+q_1q_2).
\]
Solving $\imath_Z\Omega=\mathrm{d}H_T$, when restricted to $\mathcal{M}$, we obtain
\[
Z=(Z_{\textsf{q}_1},Z_{\textsf{q}_2},Z_{\textsf{q}_3},Z_{\textsf{p}_1},Z_{\textsf{p}_2},Z_{\textsf{p}_3})=(u_1,0,u_2,-q_2u_2,-q_1u_2,0)\,.
\]
		
The consistency conditions guarantying that $Z$ is tangent to $\mathcal{M}$ are
\begin{align*}
  &0=\imath_Z\mathrm{d}p_1=Z_{\textsf{p}_1}=-q_2u_2\,,\\
  &0=\imath_Z\mathrm{d}(p_3+q_1q_2)=Z_{\textsf{p}_3}+q_2Z_{\textsf{q}_1}+q_1Z_{\textsf{q}_2}=q_2u_1\,.
\end{align*}
Therefore in the sector $q_2\neq0$ we have $u_1=u_2=0$ on $\mathcal{M}$, while in the sector defined by $q_2=0$, there are no conditions on $u_1$ and $u_2$ which are, hence, arbitrary.

Notice that the components $Z_{\textsf{q}_1}$, $Z_{\textsf{q}_2}$, $Z_{\textsf{q}_3}$ and $Z_{\textsf{p}_2}$  are just the ones that we found above and the two remaining ones just guarantee that the field $Z$ is tangent to the ``primary constraint hypersurface'' $\mathcal{M}\subset T^*\mathbb{R}^3$.

%
%
\section{Variations in \texorpdfstring{$\bm{\textrm{Emb}(\Sigma,M)}$}{Emb}}\label{appendix2}

The full explicit form of the expressions \eqref{var} is the following
\begin{align*}
&\hspace*{-3mm}Z_{\textsf{p}^\perp}=0\,,\\
&\hspace*{-3mm}Z_{\textsf{P}\alpha}=
              n_\alpha
              \Big[
                  \sqrt{\gamma}\,\gamma^{ab}\gamma^{cd} (\nabla_a Z_{\textsf{q} c})(\mathrm{d}q)_{bd}
                   -\varepsilon \gamma_{ab}\frac{Z_{\textsf{p}}^a}{\sqrt{\gamma}}p^b
                   +q_\perp \nabla_a Z^a_{\textsf{p}}
                   +Z_{\textsf{q}_\perp}\nabla_a p^a\\
                   &\hspace{8mm}
                   +\frac{1}{4}\sqrt{\gamma}\big(\nabla_e Z_{\textsf{X}}^{\top e}\!-\!Z_{\textsf{X}}^\perp K\big)\gamma^{ab}\gamma^{cd}(\mathrm{d}q)_{ac}(\mathrm{d}q)_{bd}
                   +\!\sqrt{\gamma}(Z_{\textsf{X}}^\perp K^{ab}\!-\!\nabla^a Z_{\textsf{X}}^{\top b})\gamma^{cd}(\mathrm{d}q)_{ac}(\mathrm{d}q)_{bd}\\
                   &\hspace{8mm}
                   +\frac{1}{2}\varepsilon\sqrt{\gamma} (\nabla_c Z_{\textsf{X}}^{\top c}-Z_{\textsf{X}}^\perp K)\gamma_{ab}p^a\frac{p^b}{\sqrt{\gamma}}
                   -\varepsilon \sqrt{\gamma}(\gamma_{ac}\nabla_b Z_{\textsf{X}}^{\top c}-Z_{\textsf{X}}^\perp K_{ab})p^a\frac{p^b}{\sqrt{\gamma}}\\
                   &\hspace{8mm}
                   +\varepsilon p^a (\mathrm{d}q)_{ab} (K^b_c Z^{\top c}+\varepsilon \gamma^{bc} \nabla_c Z^\perp)
                   +\varepsilon (K^b_cZ_{\textsf{X}}^{\top c}+\varepsilon \gamma^{bc}\nabla_c Z_{\textsf{X}}^\perp)q_b\nabla_a p^a
               \Big]\\
              &+e^f_\alpha
               \Big[
                    Z^a_{\textsf{p}}(\mathrm{d}q)_{af}
                    +p^a(\nabla_a Z_{\textsf{q}f}-\nabla_f Z_{\textsf{q}a})
                    +Z_{\textsf{q}f}\nabla_a p^a
                    +q_f\nabla_a Z^a_{\textsf{p}}\\
                    &\hspace{8mm}
                     -\frac{1}{4}\sqrt{\gamma}\gamma^{ab}\gamma^{cd}(\mathrm{d}q)_{ac}(\mathrm{d}q)_{bd} \big(K_{fe} Z_{\textsf{X}}^{\top e}+\varepsilon \nabla_f Z_{\textsf{X}}^\perp\big)
                     +\frac{1}{2}\varepsilon \gamma_{ab}p^a\frac{p^b}{\sqrt{\gamma}}\Big(K_{fe}Z_{\textsf{X}}^{\top e}+\varepsilon \nabla_f Z_{\textsf{X}}^\perp\Big) \\
                    &\hspace{8mm}
                     - q_\perp \Big(K_{fe}Z_{\textsf{X}}^{\top e}+\varepsilon \nabla_f Z_{\textsf{X}}^\perp\Big)\nabla_a p^a
                     -p^a(\mathrm{d}q)_{ab}(\nabla_f Z^{\top b}-Z^\perp K^b_f)\\
                     &\hspace{8mm}
                     -(\nabla_f Z_{\textsf{X}}^{\top b}-Z_{\textsf{X}}^\perp K^b_f)q_b\nabla_a p^a
                \Big]\,.
\end{align*}
In order to compute them, the following variations of geometric objects on $\mathrm{Emb}(\Sigma,M)$ are helpful
\begin{align}
&D_{Z_{\textsf{X}}}e^a_\alpha=\varepsilon n_\alpha\left(K^a_b Z_{\textsf{X}}^{\top b}+\varepsilon\gamma^{ab}(\mathrm{d}Z_{\textsf{X}}^\perp)_b\right)-e^b_\alpha \left(\nabla_b Z_{\textsf{X}}^{\top a}-Z_{\textsf{X}}^\perp K_b^a\right)\,,\\
&D_{Z_{\textsf{X}}}n_\alpha=-e_\alpha^a\Big(K_{ab}Z_{\textsf{X}}^{\top b}+\varepsilon (\mathrm{d}Z_{\textsf{X}}^\perp)_a \Big)\,,\\
&D_{Z_{\textsf{X}}}\gamma_{ab}=\gamma_{bc}\nabla_a Z_{\textsf{X}}^{\top c}+\gamma_{ac}\nabla_b Z_{\textsf{X}}^{\top c}-2Z_{\textsf{X}}^\perp K_{ab}\,,\\
&D_{Z_{\textsf{X}}}\gamma^{ab}=-\nabla^aZ_{\textsf{X}}^{\top b}-\nabla^bZ_{\textsf{X}}^{\top a}+2Z_{\textsf{X}}^\perp K^{ab}\,,\\
&D_{Z_{\textsf{X}}}\sqrt{\gamma}=\Big(\nabla_aZ_{\textsf{X}}^{\top a}-Z_{\textsf{X}}^\perp K_a^a\Big)\sqrt{\gamma}\,,\\
&D_{Z_{\textsf{X}}}(\nabla_ap^a)=0\,.
\end{align}
Here  $K_a^b$ denotes the Weingarten map associated with $X(\Sigma)\subset M$, indices are lowered and raised with $\gamma_{ab}$ and $\gamma^{ab}$ respectively, and $\nabla$ denotes the Levi-Civita connection of $(\Sigma,\gamma)$ (which depends on the embedding $X$ through $\gamma$). Notice that as $p^a$ is a vector density $\nabla_ap^a$ depends only on the differential structure on $\Sigma$ (i.e.\ it does not matter what connection we use to define it); this is why we get zero when computing $D_{Z_{\textsf{X}}}$ acting on it so that $D_Z(\nabla_ap^a)=\nabla_aZ_{\textsf{p}}^a$. These expressions and their derivations are discussed in \cite{Kuchar1,Kuchar2,Kuchar3} and, from a slightly different perspective in \cite{Bauer}.

\end{appendices}

%
%

\end{document}